\documentclass[a4paper,10pt,twoside]{cpc-hepnp}

\usepackage{multicol}
\usepackage{graphicx}
\usepackage{booktabs}
\usepackage{amssymb,bm,mathrsfs,bbm,amscd}
\usepackage[tbtags]{amsmath}
\usepackage{lastpage}

\begin{document}

\fancyhead[c]{\small Submitted to 'Chinese Physics C'}

\title{High-spin level structure of the neutron-rich nucleus $^{91}$Y}

\author{Xiao-Feng HE$^{1,2;1}$\email{hexiaofeng@impcas.ac.cn}, Xiao-Hong ZHOU$^{1;2}$\email{zxh@impcas.ac.cn},
Yong-De Fang$^1$, Min-Liang Liu$^1$, Yu-Hu ZHANG$^1$, Kai-Long WANG$^{1,2}$, \\
Jian-Guo WANG$^1$, Song GUO$^1$, Yun-Hua QIANG$^1$, Yong ZHENG$^1$, Ning-Tao ZHANG$^1$, Guang-Shun LI$^1$, \\
Bing-Shui GAO$^{1,2}$, Xiao-Guang WU$^3$, Chuang-Ye HE$^3$, Yun ZHENG$^3$ \\
}
\maketitle

\address{%
$^1$ Key Laboratory of High Precision Nuclear Spectroscopy and Center for Nuclear Matter Science,
Institute of Modern Physics, \\
Chinese Academy of Sciences, Lanzhou 730000, People's Republic of China \\
$^2$ University of Chinese Academy of Sciences, Beijing 000049, People's Republic of China \\
$^3$ China Institute of Atomic Energy, Beijing 102413, People's Republic of China \\
}

\begin{abstract}
High-spin level structure of the neutron-rich nucleus $^{91}$Y has been reinvestigated via the
$^{82}$Se($^{13}$C, $p3n$)$^{91}$Y reaction. A newly constructed level scheme including several key levels clarifies the
uncertainties in the earlier studies. These levels are characterized by the
breaking of the $Z=38$ and $N=56$ subshell closures, which involves in the spin-isospin
dependent central force and tensor force.
\end{abstract}



\begin{multicols}{2}

\section{Introduction}

It is well known that the high-spin states of near-spherical nuclei can be
constructed by the aligned angular momentum of open shell nucleons.
The maximum spin occurs at the configuration termination in the
valence space. To generate higher-spin states, the shell closure should be broken.
A number of studies have revealed such excitation process for nuclei
around the quasidoubly magic nucleus $^{88}$Sr
\cite{funke1992,pattabiraman2002,hausmann2003,jungclaus1999}.

The low-lying levels of nuclei with $38<Z<50$ and $50<N<56$ are dominated
by the valence-nucleon excitations within the
$\pi (p_{1/2}, g_{9/2})$ and $\nu d_{5/2}$ shell-model orbitals. The medium-
or high-spin levels can be understood as the particle-hole excitations. Federman \emph{et al.}
have theoretically studied the proton $ p_{3/2} \rightarrow p_{1/2}$ and neutron
$d_{5/2} \rightarrow g_{7/2}$ excitations \cite{federman1977,federman1984}
and given the reasonable explanations for the reductions of
the $Z=38$ and $N=56$ energy gaps. The proton-neutron interaction
includes tensor force and the spin-isospin
dependent central force \cite{otsuka2001,otsuka2005}. In the picture of tensor
force \cite{otsuka2001}, the enhanced $\pi p_{1/2}-\nu d_{5/2}$ and reduced $\pi p_{3/2}-\nu d_{5/2}$
monopole interaction may arise from the tensor component if the others
contribute the same interactional values.
Hence, when the $\nu d_{5/2}$ orbital is occupied by more neutrons, the $\pi p_{1/2}$
orbital goes down while $\pi p_{3/2}$ orbital comes up, that is, the $Z=38$ energy gap becomes smaller.
Likewise, the reduction of the $N=56$ gap can be associated with the spin-orbital
partners $\pi g_{9/2}$ and $\nu g_{7/2}$  \cite{federman1984}, between which
the interaction is the so-called spin-isospin dependent central force \cite{otsuka2005}.

The production of neutron-rich nucleus $^{91}$Y is problematic via the conventional ($HI$, $x$n) reaction.
High-spin states of $^{91}$Y have been initially investigated via fusion-evaporation reaction
$^{82}$Se($^{12}$C, p2n)$^{91}$Y \cite{Bucurescu}.
It was not until recently that an extended level scheme is constructed by means of the
fission of $^{221}$Pa produced by a $^{24}$Mg beam impinging on a
$^{173}$Yb target \cite{Fotiades}. In this work, the high-spin level
structure is reinvestigated and the configurations are figured out,
in analogy to our earlier study of the $N=52$ isotone $^{92}$Zr \cite{Wang}.

\section{Experiment}

High-spin states of $^{91}$Y were populated
through the $^{82}$Se($^{13}$C, $p3n$)$^{91}$Y reaction. The $^{13}$C
beam was provided by the HI-13 Tandem
Accelerator of the China Institute of Atomic Energy(CIAE), and the
target was 2.11 mg/cm$^{2}$ isotopically
enriched $^{82}$Se on 8.56 mg/cm$^{2}$ natural lead backing. The
emitted $\gamma$ rays from the reaction products
were detected by an array consisting of 2 planar and 12 Compton-suppressed
HPGe detectors. The energy and efficiency
calibrations were made using the $^{60}$Co, $^{133}$Ba, and $^{152}$Eu
standard sources and the typical energy
resolution was 2.0$\sim $2.5 keV at full width at half-maximum (FWHM)
for the 1332.5-keV $\gamma$ ray of $^{60}$Co.
Events were collected when at least 2 detectors are fired within the
prompt 80 ns coincidence time window. Under
these conditions, a total of 2.5$\times 10^{7}$ coincidence events
were recorded and the data was sorted into a
symmetrized $E_{\gamma}-E_{\gamma}$ matrix for subsequent off-line
analysis.

In order to obtain multipolarity information of the emitted $\gamma$ rays,
two asymmetric coincidence matrices were
constructed using the $\gamma$ rays detected at all angles (as y axis)
against those observed at $\sim 34^\circ$ (or
$\sim 146^\circ$) and 90$^\circ$ angles (as x axis) respectively. The ADO
($\gamma$ ray angular distribution from oriented nuclei)
ratio \cite{piiparinen1996} is defined as
$R_{\rm ADO}(\gamma)=\frac{I_\gamma(34^\circ)}{I_\gamma(90^\circ)}
=\frac{N_\gamma(34^\circ)/\epsilon_\gamma(34^\circ)}{N_\gamma(90^\circ)/\epsilon_\gamma(90^\circ)}$
(here $\gamma$ refers to a particular energy value.). By
carefully setting gates on the y-axis with all the detectors, the
$\gamma$ ray intensities $I_\gamma (34^\circ)$ and
$I_\gamma (90^\circ)$ were extracted from the coincidence spectra
regardless of the multipole character of the gating
transition. Generally, one transition is adopted as stretched quadrupole transition
if its $R_{\rm ADO}(\gamma)$ is significantly
larger than 1.0 and it's adopted as dipole transition if its
$R_{\rm ADO}(\gamma)$ is less than 1.0.

\section{Experimental Results}

We have noticed a significant difference between the level
schemes given in the earlier studies \cite{Bucurescu,Fotiades}. To the extent of our
knowledge, it is difficult to understand
the identified strong $\gamma$ rays in Ref.~\cite{Fotiades}
not listed in Ref.~\cite{Bucurescu}. In order to clarify the uncertainties,
we have carefully checked the $\gamma$-$\gamma$ coincidence relationships and
construct a new level scheme shown in Fig.~\ref{fig1}. Except for 3569-,
5579.9-, 5781.2-keV levels and 1411.6-, 953.5-, 1094.5-, 1297.8-keV
$\gamma$ rays (see Fig.~\ref{fig2} for $\gamma$-$\gamma$ coincidence
relationships), the other levels and $\gamma$ rays in \cite{Fotiades}
can not be confirmed in the present work; it is worth pointing out that the
766.8- and 579.3-keV $\gamma$ rays are weak in Fig.~\ref{fig2} but can be confirmed in coincidence with the 672.3-, 335.0- and
327.7-keV lines.

On referring to the prior studies \cite{Bucurescu,Fotiades}, an obvious contribution of our
work is the revisional or tentative assignment of the spin-parity values
of the 3734.4-, 4483.4- and 4811.1-keV levels. According to the
ADO ratios of $\gamma$ rays listed in Table~\ref{tab1},
the 1577-keV $\gamma$ ray is a quadrupole transition and the spin of
the 3734.4-keV level is therefore $(21/2)$. Accepting the spins and parities
of 3528.4- and 4148.4-keV levels in \cite{Bucurescu}, the intensity
ratio between the 415.3- and 619.6-keV lines
proposes that the former is an E2 rather than M2 transition and the
parity of the 3734.4-keV level is therefore positive. It can be seen from
Fig.~\ref{fig3} that the 327.7- and 335-keV lines are obviously dipole
transitions if the 619.6-, 929.5-, 1371.0- and 1577.0-keV lines are quadrupole
ones. Hence, the 4483.4-keV level is $(27/2)$. Considering the
intensity ratio of the 335.0- and 953.5-keV $\gamma$ rays, the 953.5-keV $\gamma$ ray
should be an E3 rather than M3 transition and the parity of the 4483.4-keV
level is of course negative. $(29/2)$ is tentatively adopted for the
4811.1-keV level according to the dipole character of the 327.7-keV line.

\begin{center}
\includegraphics[width=6cm]{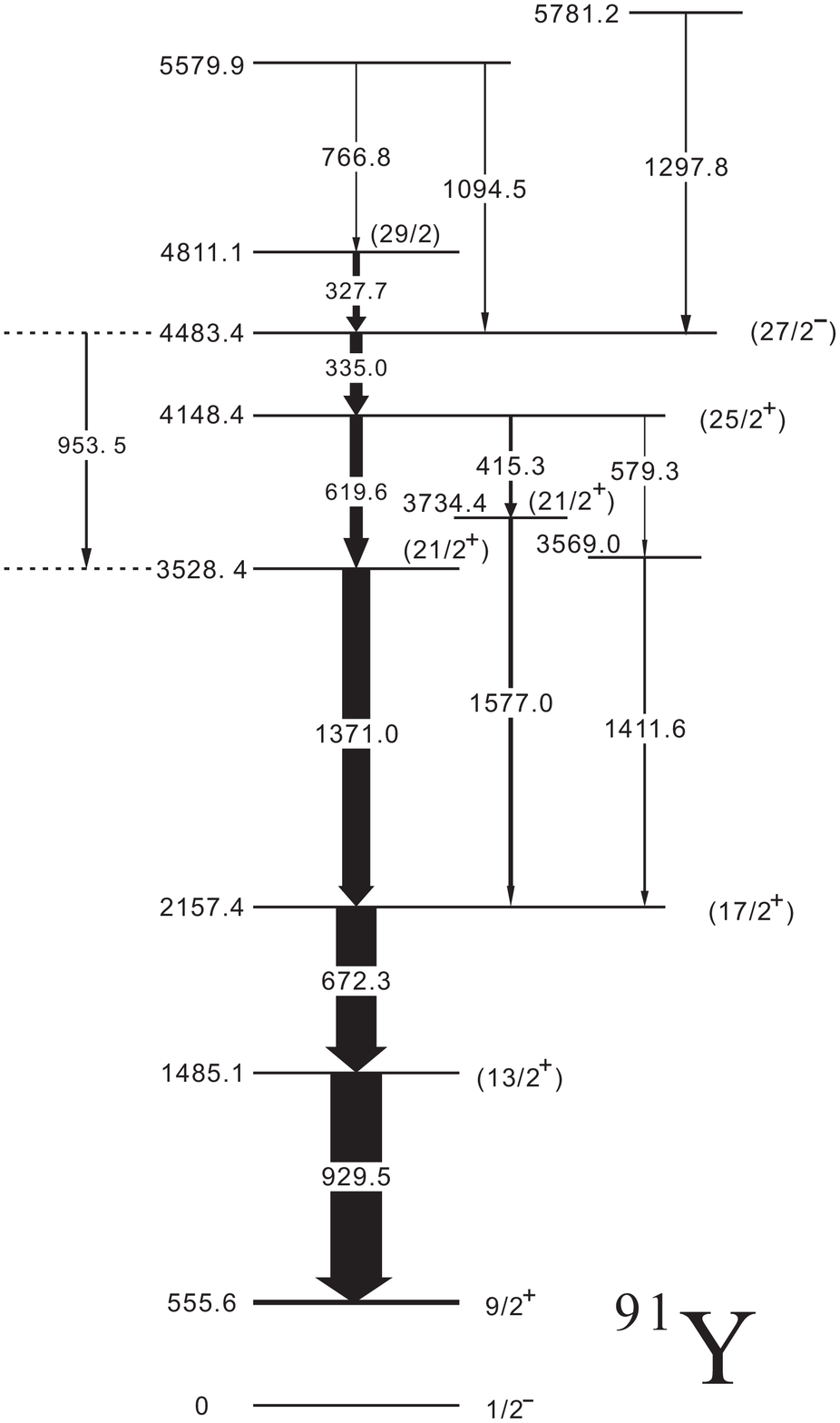}
\figcaption{The revised level scheme of $^{91}$Y deduced from the
present work. The 579.3-, 1411.9-,
953.5-, 1297.8-, 766.8- and 1094.5-keV $\gamma$ rays are newly added into the level
scheme compared with that in Ref.~\cite{Bucurescu}. }
\label{fig1}
\end{center}

\begin{center}
\includegraphics[width=8cm]{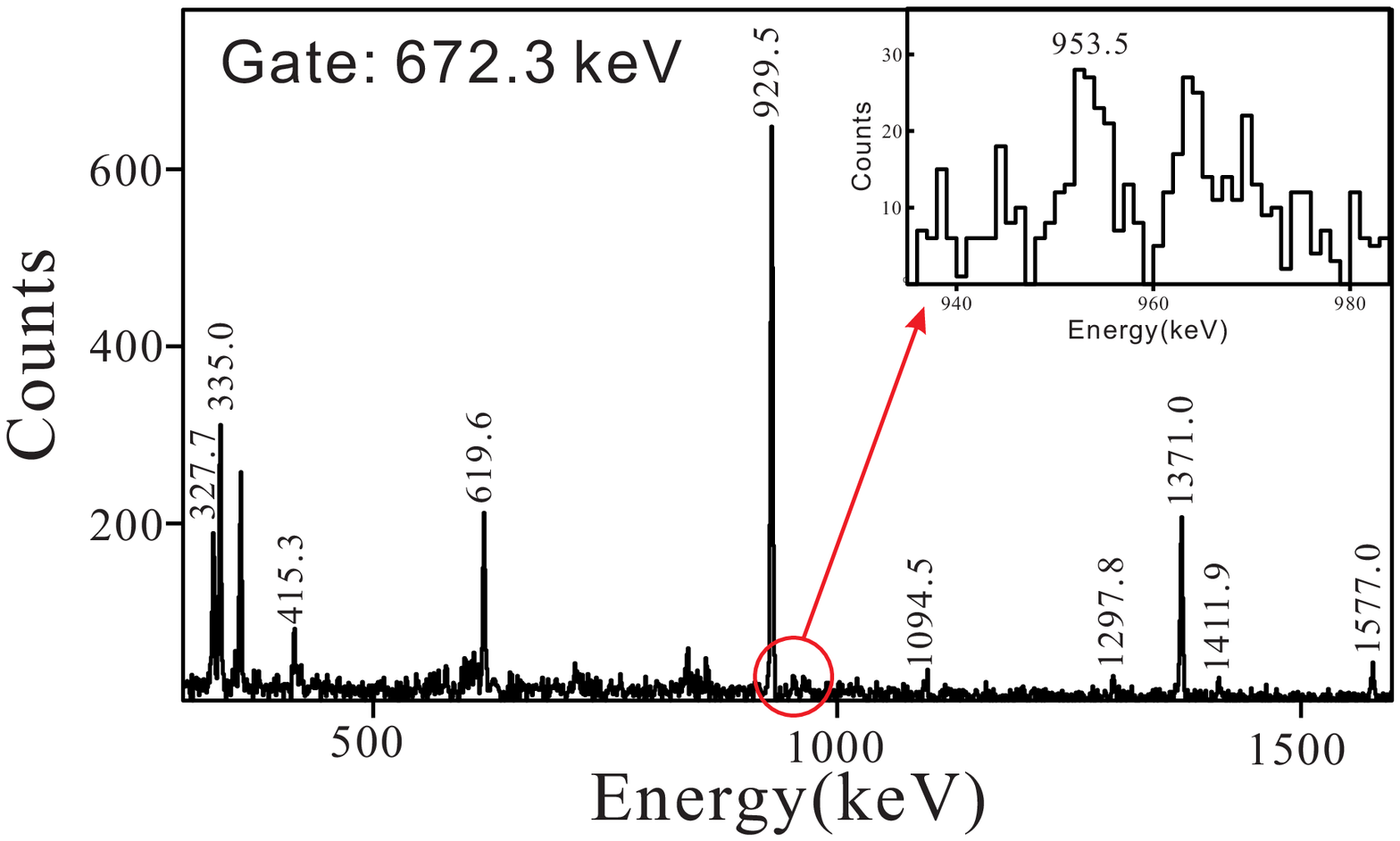}
\figcaption{Typical coincidence spectrum for $^{91}$Y gated on the 672.3-keV line.}
\label{fig2}
\end{center}

\begin{center}
\includegraphics[width=8cm]{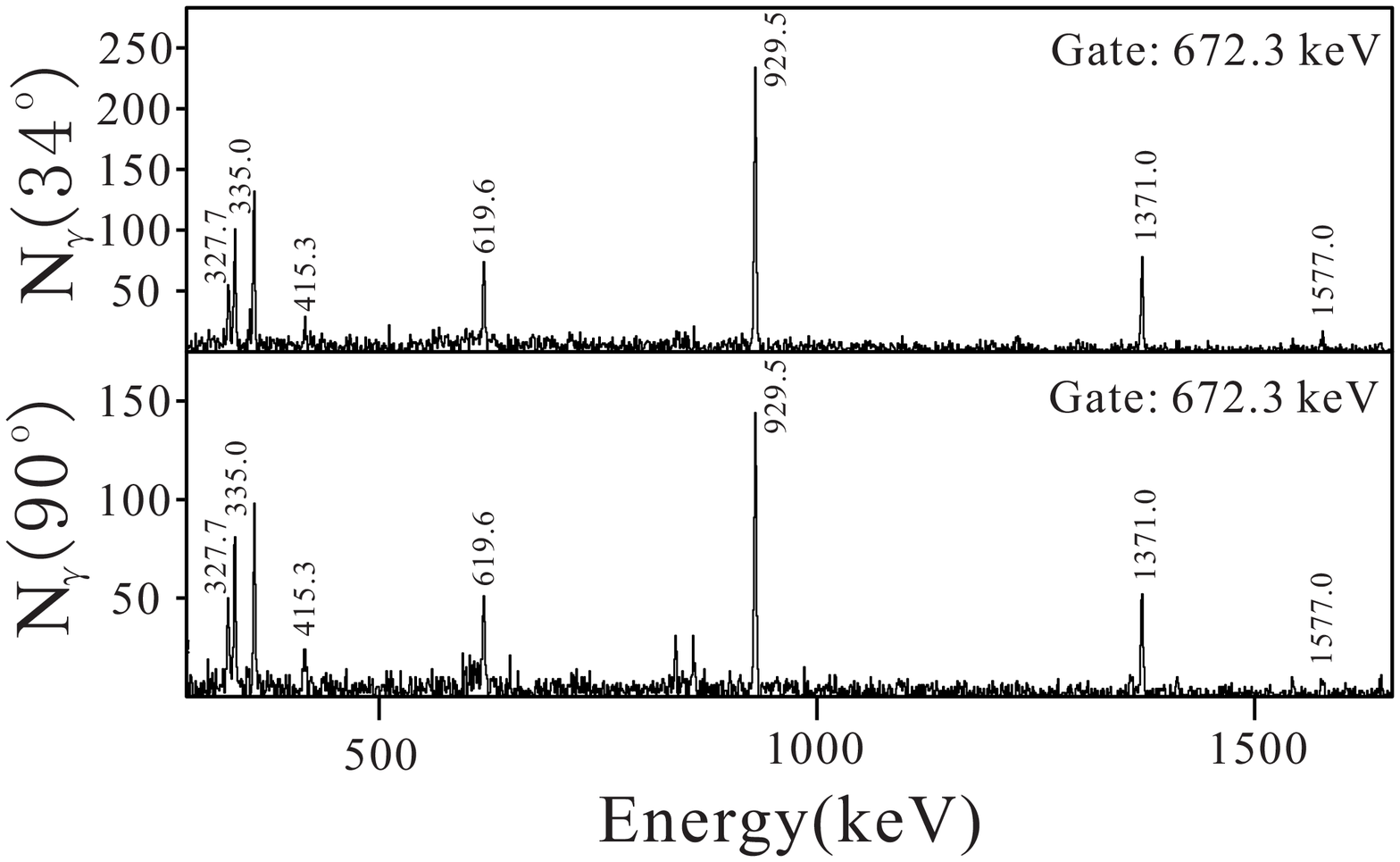}
\figcaption{$\gamma$ ray counts $N_\gamma(34^\circ)$ and $N_\gamma(90^\circ)$ of
the two asymmetry matrices at $34^\circ$ and $90^\circ$ angles respectively gated on
the 672-keV line. It's very clear that the 327.7- and 335.0-keV
 $\gamma$ rays are dipole transitions for
their ADO ratios are smaller than 1.0 while the 619.6-, 929.5-, 1371.0- and 1577.0-keV lines are
quadrupole transitions for their ADO ratios are larger than 1.0. }
\label{fig3}
\end{center}

\begin{table*}[p]
\centering
\caption{The $\gamma$ ray transition energies, relative intensities, ADO ratios, and their assignments in $^{91}$Y.}  \vspace{0.5mm}
\label{tab1}
\begin{tabular}{lllll}
\hline
$E_{\gamma}$(keV) $^{a}$ & $I_{\gamma}$ $^{b}$ & $R_{ADO}$ & $E_{i}  \rightarrow E_{f}$(keV) $^{c}$ & $J^{\pi}_{i} \rightarrow J^{\pi}_{f}$ $^{d}$ \\
\hline
327.7                    & 17.7                & 0.42(5)   & $4811.2 \rightarrow 4483.5$            & $(29/2)      \rightarrow (27/2^-)$ \\
335.0                    & 31.0                & 0.46(7)   & $4483.5 \rightarrow 4148.5$            & $(27/2^-)    \rightarrow (25/2^+)$ \\
415.3                    & 9.1                 & 1.45(19)  & $4148.5 \rightarrow 3734.3$            & $(25/2^+)    \rightarrow (21/2^+)$ \\
579.3                    & 7.1                 &           & $4148.5 \rightarrow 3569.7$            & $(25/2^+)    \rightarrow         $ \\
619.6                    & 32.0                & 1.06(15)  & $4148.5 \rightarrow 3528.4$            & $(25/2^+)    \rightarrow (21/2^+)$ \\
672.3                    & 100                 & 1.29(13)  & $2157.4 \rightarrow 1485.1$            & $(17/2^+)    \rightarrow (13/2^+)$ \\
766.8                    & 3.8                 &           & $5578.0 \rightarrow 4811.2$            & $            \rightarrow (29/2)  $ \\
929.5                    & $\ge$127.6          & 1.41(13)  & $1485.1 \rightarrow 555.6 $            & $(13/2^+)    \rightarrow 9/2^+   $ \\
953.0                    & 6.6                 &           & $2131.1 \rightarrow 1882.4$            & $(27/2^-)    \rightarrow (21/2^+)$ \\
1094.5                   & 5.0                 &           & $5578.0 \rightarrow 4483.5$            & $            \rightarrow (27/2^-)$ \\
1297.8                   & 4.9                 &           & $5781.3 \rightarrow 4483.5$            & $            \rightarrow (27/2^-)$ \\
1371.0                   & 69.3                & 1.27(23)  & $3528.4 \rightarrow 2157.4$            & $(21/2^+)    \rightarrow (17/2^+)$ \\
1411.6                   & 6.0                 &           & $3569.7 \rightarrow 2157.4$            & $            \rightarrow (17/2^+)$ \\
1577.0                   & 11.1                & 1.30(43)  & $3734.3 \rightarrow 2157.4$            & $(21/2^+)    \rightarrow (17/2^+)$ \\
\hline
\end{tabular}
\flushleft
{$^{a}$ Uncertainties are within 0.5 keV.} \\
{$^{a}$ Uncertainties are within 30\%.} \\
{$^{c}$ Excitation energies of initial $E_{i}$ and final $E_{f}$ states.} \\
{$^{d}$ Proposed spin and parity assignments for the initial $J^{\pi}_{i}$ and final $J^{\pi}_{f}$ levels.} \\
\end{table*}

\section{Discussion}

Since the Fermi levels of $^{91}$Y lie at the $\pi p_{1/2}$ and
$\nu d_{5/2}$, the active proton and neutron orbitals are $(f_{5/2}, p_{3/2}, p_{1/2},
g_{9/2})$ and $(d_{5/2}, s_{1/2}, d_{3/2}, g_{7/2}, h_{11/2})$,
respectively (for the single-particle energies see Ref.~\cite{Wang} ).
The $9/2^+$ state is naturally the $g_{9/2}$ proton coupled with two $ d_{5/2}$ neutrons in pair.
It is expected that the $(13/2^+)$ level above the
$9/2^+$ can be associated with the $\pi g_{9/2} \otimes \nu (d^2_{5/2})_{2^+}$. The three-particle
$\pi g_{9/2} \otimes \nu d^2_{5/2}$ configuration may persist till the $(17/2^+)$.
The $(17/2^+)$ is the maximum spin for the termination of the configuration
in the space $\pi (p_{1/2}, g_{9/2})$ and $\nu d_{5/2}$.

Above the fully aligned $(17/2^+)$, the states should therefore arise
from the competition of one proton and one neutron excitations across the respective $Z=38$ $N=56$
subshell closures. It has been pointed out in our previous results \cite{Wang}
that the low-energy one is attributed to the neutron excitation. In the same way, we propose
that the excitation from the 2157.4-keV ($17/2^+$) to 3528.4-keV ($21/2^+$) level
is $\nu(d_{5/2} \rightarrow g_{7/2})$ and the 3528.4-keV state has the
$\pi g_{9/2} \otimes \nu(d_{5/2}g_{7/2})$ configuration. The $\nu g_{7/2}$
single-particle excitation energy relative to the $\nu d_{5/2}$  decreases
from $\sim$2.19 MeV in $^{91}$Zr to $\sim$0.17
MeV in $^{101}$Sn \cite{ENSDF}, namely $\sim$0.2 MeV
lowered by adding a proton into the $g_{9/2}$ orbital. The energy of the $\nu (d_{5/2} \rightarrow
g_{7/2})$ excitation in $^{92}$Zr ($(12^{+}) \rightarrow (14_1^{+})$) is 1098.4
keV. Thus we can predict
the excitation energy of the $\pi g_{9/2} \otimes \nu d^2_{5/2} \rightarrow \pi
g_{9/2} \otimes \nu (d_{5/2}g_{7/2})$ in $^{91}$Y to be $\sim$1298.4 (=1098.4+200) keV;
the 1371-keV energy (see Fig.~\ref{fig1}) agrees reasonably well with this
value. The lowered $\nu(d_{5/2} \rightarrow g_{7/2})$ excitation energy reflects the dramatic
reduction of the $N=56$ energy gap due to the spin-isospin dependent central force.

Given the 3528.4-keV state as neutron particle-hole excitation, the other
($21/2^+$) at 3734.4 keV is naturally attributed to the proton particle-hole
excitation. Since the Z=38 energy gap is separated by the $\pi p_{3/2}$ and
$\pi p_{1/2}$, the particle-hole excitation might be $\pi (p_{3/2} \rightarrow p_{1/2})$.
Thus the energy gap in $^{91}$Y (1577 keV ) is $\sim$0.233 MeV lower than
$\sim$1.8 MeV to break the $Z = 38$ subshell closure in $^{88}$Sr which is demonstrated by the
excitation energy of the $2^{+}$ state \cite{ENSDF}. This indicates a possible reduction of the energy gap when
two neutron goes into the $d_{5/2}$ orbital. From the introduction part,
the reduction may be associated with the proton-neutron tensor force \cite{otsuka2005}.
To generate the $(25/2)$ spin value, both of the $Z = 38$ and $N = 56$ subshell closures should
be broken. Therefore, the $\pi (p_{3/2}^{-1}p_{1/2}g_{9/2}) \otimes \nu
(d_{5/2}g_{7/2})$ configuration is proposed for the $(25/2^+)$ state.

For achieving higher-spin state, the proton should be further excited
to the $g_{9/2}$, namely $\pi (p_{3/2}/f_{5/2} \rightarrow g_{9/2})$.
The $(27/2^-)$ and $(29/2)$ states appear to be the terminations of the
$\pi (p_{3/2}^{-1}g^2_{9/2}) \otimes \nu (d_{5/2}g_{7/2})$ and
$\pi (f_{5/2}^{-1}g^2_{9/2}) \otimes \nu (d_{5/2}g_{7/2})$ configurations, respectively.

\section{Summary}

The high-spin level structure of $^{91}$Y has been reinvestigated through the $^{82}$Se($^{13}$C, $p3n$)$^{91}$Y
fusion-evaporation reaction. The level scheme has been modified. The spin-parity values
of states in $^{91}$Y above the fully aligned $(17/2^+)$  state have been proposed. The configuration of high-spin
states above the fully aligned $(17/2^+)$  state are figured out in terms of the breaking of the
$Z=38$ and $N=56$ subshell closures. The tensor force and spin-isospin dependent central force play
an important role in breaking the subshells.

\section{Acknowledgments}

We would like to thank the tandem accelerator group at the China Institute
of Atomic Energy for their support during the experiments. This work has
been supported by the Major State Basic Research Development Program of China (Grant
No. 2013CB834403), the National Natural Sciences Foundation (Grants Nos. 11175217, 11205207, 11205208, and U1232124), and the Chinese Academy of Sciences.

\end{multicols}

\clearpage

\end{document}